# Rediscovering Black Phosphorus: A Unique Anisotropic 2D Material for Optoelectronics and Electronics

Fengnian Xia[1], Han Wang[2], and Yichen Jia[1]

[1]*Department of Electrical Engineering, Yale University, New Haven, Connecticut 06511*

[2]*IBM Thomas J. Watson Research Center, Yorktown Heights, NY 10598*

**Abstract.** Anisotropy refers to the property of a material exhibiting directionally dependent features. In this paper, we introduce black phosphorous (BP), the most stable allotrope of phosphorus in layered orthorhombic structure with a bandgap of 0.3 eV in bulk, as a unique 2D material in which electrons, phonons and their interactions with photons behave in a highly anisotropic manner within the plane of the layers. The unique anisotropic nature of BP thin films is revealed using angle-resolved Raman and infrared spectroscopies, together with angle-resolved transport study. For 15 nm thick BP, we measure Hall mobility of 1000 and 600 $cm^2/Vs$ for holes along the light (x) and heavy (y) effective mass directions, respectively, at 120 K. These BP thin films also exhibit large and anisotropic in-plane optical conductivity from 2 to 5 μm wavelength. Field effect transistors using 4 to 30 layers of BP (2 to 15 nm) as channel material exhibit an on-off current ratio exceeding $10^5$, a field-effect mobility of 205 $cm^2/Vs$, and good saturation properties all at room temperature, suggesting its promising future in high performance thin film electronics. By introducing narrow bandgap BP into the 2D material family, we fill the space between semi-metallic graphene and large bandgap TMDCs, where great potentials for infrared optoelectronics lie. Most importantly, the unique anisotropic nature of this intriguing material creates unprecedented possibilities for the realization of conceptually new optoelectronic and electronic devices in which angle-dependent physical properties are highly desirable.

Current research in two-dimensional materials primarily focuses on graphene[1-10], the insulating hexagonal boron nitride (hBN)[11-13], and members from the transition metal dichalcogenides (TMDCs)[14-21] such as molybdenum disulfide ($MoS_2$) and tungsten diselenide ($WSe_2$). However, graphene suffers from limitations for electronic and photonic applications due to its zero-bandgap nature while most of the TMDCs currently being studied have band gaps corresponding to the visible spectrum range, rendering them unsuitable for optoelectronics at the infrared frequencies where greater technological demands exist. Although transistors made from few-layer TMDCs such as $MoS_2$ and $WSe_2$ show very high on-off current ratios and excellent current saturation properties, due to their sizable bandgap larger than 1.5 eV, the carrier mobility in these TMDCs is much lower than in graphene and fluctuates significantly from less than 10 to around 200 $cm^2/Vs$[21, 22-23], restricting their applications in both optoelectronics and electronics. Furthermore, in these 2D materials, the photonic, electronic, and mechanical properties within the plane of the layers are largely isotropic and do not exhibit significant directional dependence. In fact, the concept of utilizing strong anisotropic properties of 2D materials for novel optoelectronic and electronic device applications has never been proposed before, nor is it possible using any previously studied 2D materials.

In this work, we introduce black phosphorus (BP) to the 2D material family as a unique anisotropic material for optoelectronic and electronic applications. Narrow gap BP can fill the space between zero gap graphene and large gap TMDCs, making it an ideal 2D material for near and mid-infrared optoelectronics. BP thin films show high mobility above 600 $cm^2/Vs$ at room temperature and above 1000 $cm^2/Vs$ at 120 K along the light effective mass (x) direction. Transistors made from BP (2 to 15 nm) demonstrate on-off current ratio above $10^5$, on-state current exceeding 200 µA/µm, and good saturation properties, implying its promising future for high frequency, thin-film electronics. Most importantly, the unique anisotropic nature within the plane of layers may allow for the realization of novel optoelectronic, electronic, and nano-mechanical devices.

1. **Characterization of BP thin film layer number**

Similar to graphite, bulk black phosphorus (BP) is an elemental layered material. Bulk BP crystal has been studied extensively decades ago and it is the most stable allotrope of phosphorus[24-28], due to its unique orthorhombic crystal structure as shown in Figure 1a. As carbon atoms in graphene, each phosphorus atom is connected to three adjacent ones in black phosphorus to form a stable, linked ring structure, each consisting of six phosphorus atoms. However, the puckered BP structure has reduced symmetry compared to graphite, resulting in its unique angle-dependent in-plane conductivities[29-32]. The adjacent layer spacing is around 0.53 nm and the lattice constant in orthorhombic system along z-direction is 1.05 nm. Moreover, as in TMDCs, the bandgap in BP is expected to be strongly dependent on the layer number[33], due to the layer-layer coupling. Recently, the dependence of its band structure on strain was also investigated theoretically[34].

In our experiments, few-layer BP was first exfoliated using the standard mechanical cleavage method from bulk BP crystal and then deposited on silicon substrate with 300 nm silicon dioxide. A typical atomic force microscope (AFM) image of few-layer BP is shown in Fig. 1b. Using a layer-to-layer spacing of 0.53 nm[32], we clearly identified few-layer BP with thickness down to around 1.2 nm, corresponding to 2 atomic layers of BP. However, these few-micrometer scale bilayer BP flakes are usually too small for optical and electronic experiments. In this work, we performed optical and electrical measurements on BP layers with a thickness varying from 2 to > 30 nanometers (4 to 60 layers).

## 2. Angle-resolved IR spectroscopy and DC conductance

For BP films thicker than 8 nm, flakes with lateral size in the tens of micrometers scale can be isolated, allowing for the determination of crystalline directions using infrared extinction spectroscopy[35,36]. The lower right inset of Fig. 2a shows an optical image of a BP thin film with a thickness of around 30 nanometers. We measured the angle-resolved optical extinction spectrum (1-T/$T_0$), where T and $T_0$ are transmission through the BP and the reference, respectively. Figure 2a denotes the extinction spectra measured on BP with the incident light from the z-direction at six different linear polarization angles separated by 30 degrees, indicated by the six arrowed lines with different colors in the inset. In this experiment, the 0° reference direction was selected arbitrarily. For all polarizations, the extinction shows a sharp increase at around 2400 cm$^{-1}$, indicating a band gap of around 0.3 eV, which agrees very well with previously reported BP band gap at Z symmetry point[32]. Since the optical conductivity of BP peaks at the x-direction at the band edge[32], by rotating the linear polarizer in finer steps, the angle between the x-direction and the blue arrow (0° in Fig. 2a) is determined to be -8 degrees. The anisotropy in optical conductivity arises from the directional dependence of the interband transition strength in anisotropic BP bands[32,37].

We also independently determined the x- and y-directions of the BP thin film using angle-resolved DC conductance measurement. In this experiment, 12 electrodes (1 nm Ti/20 nm Pd/30 nm Au) were fabricated on the same flake spaced at an angle of 30 degrees along the directions as shown in the inset of Fig. 2b, which are along the same directions indicated by the color lines in Fig 2a. Here, we use the same 0° reference as that in the angle-resolved IR spectroscopy measurement in Fig. 2a. We performed the DC conductance measurements by applying electric field across each pair of diagonally positioned electrodes separated by 22 μm at 180 degrees apart and the results are plotted in Fig. 2b in polar coordinates. Measurements on each pair of electrodes lead to two data points in Fig. 2b separated by 180 degrees for positive and negative bias, respectively. The directional dependence of low-field conductivity in anisotropic material can be described by the equation: $\sigma_\theta = \sigma_x \cos^2(\theta-\phi) + \sigma_y \sin^2(\theta-\phi)$. The resulting calculated conductance fits nicely to the measurement data (red solid curve in Fig. 2(b)). $\sigma_x$ and $\sigma_y$ refer to the conductivity of BP along x- and y-directions respectively. $\theta$ is the angle, with respect to the 0° reference direction, along which the electric field is applied and the conductance is measured. $\sigma_\theta$ is the conductivity along $\theta$ direction. $\phi$ is the angle between the x-direction and the 0° reference. $\phi = -8$ degree gives the best fit for this sample (Fig. 2b), which agrees perfectly with the value obtained using IR extinction spectroscopy. Here, the x- and y-directions correspond to the high and low mobility directions as shown in Fig. 1. From the fitted curve, the ratio $\sigma_x / \sigma_y$ is extracted to be around 1.5, corresponding to a ratio of mobility along x- and y-directions $\mu_x/\mu_y$ of 1.5. Due to the current spreading, this simple approach probably leads to underestimation of the mobility contrast, as confirmed by the Hall measurement presented in the later part of this work.

3. **Angle-resolved Raman scattering**

We further performed angle-resolved Raman scattering measurements on the same BP flake discussed in sections 1 and 2 with linearly polarized laser excitation incident from the z-direction. This experimental configuration (z in, z out) is the most relevant for layered BP thin films. Previous IR and DC conductivity experiments allow us to identify the crystalline orientation of the BP flake. The Raman scattering results for excitation light polarizations along x, D, and y directions are plotted in Fig. 3. Here, D-direction is along 45° to both x and y directions. We observe three Raman peaks in thin film BP at around 470, 440, and 365 cm$^{-1}$, corresponding to $A_g^2$, $B_{2g}$, and $A_g^1$ modes regardless of the polarization, which agrees well with previous observations in bulk BP crystal[38]. Three other Raman active modes are not observed here due to the selection rules in this particular experimental configuration (z in, z out). For all three different polarizations, we did not observe measurable peak position variations as polarization changes. However, the relative intensity of these three peaks does change significantly. For example, as show in Fig. 3, $B_{2g}$ clearly peaks when incident light polarization is along D-direction, because its strength is determined only by off-diagonal polarizability tensor $\alpha_{xy}$ and polarization at 45° maximizes the strength[39]. $A_g^2$ mode only shows insignificant intensity variation according to the polarization. On the contrary, the intensity of $A_g^1$ mode varies significantly when polarization changes. For both $A_g$ modes, the Raman scattering intensity is determined by both diagonal and off-diagonal elements of polarizability tensor and detailed theoretical investigations are needed to account for these variations. For few-layer BP film down to 3 nm, we also observed the same polarization dependence in the relative intensity of $A_g^2$, $B_{2g}$, and $A_g^1$ modes, but again did not observe any significant frequency shift in these three modes. While a more detailed discussion and modeling of the (z in, z out) angle-resolved Raman spectroscopy and its dependence on the thin film thickness is not the focus of this paper, the angle-resolved Raman measurements here offers another method for determining the crystalline directions of the BP thin film crystals.

### 4. Angle-resolved Hall mobility

Using IR extinction spectroscopy method described in section 2, we first identified crystalline directions in BP thin film with a thickness of 8 and 15 nanometers. Standard Hall bars were fabricated on these thin films and the mobilities were measured at both x- and y-directions from 10 Kelvin to room temperature, as shown in Fig. 4. For the 15 nm sample, the mobility along x-direction exceeds 600 $cm^2$/Vs at room temperature and is above 1000 $cm^2$/Vs at 120 K. For the 8 nm sample, the mobility along x-direction exceeds 400 $cm^2$/Vs at room temperature and is above 600 $cm^2$/Vs at 120 K. The hole carrier density is kept at $6.7 \times 10^{12}$ $cm^{-2}$ in all the measurements above. For both thicknesses, the Hall mobility along x-direction is about 1.8 times as large as that in y-direction, representing anisotropy slightly smaller than that in bulk BP[32], but slightly larger than that extracted from the DC conductance measurement in Fig. 2b. Thicker BP consistently shows higher mobility, indicating the impact of the substrate on the scattering of the carriers in BP thin film[40]. In bulk BP crystals, hole mobility above 2000 $cm^2$/Vs and 50,000 $cm^2$/Vs along x-direction were previously reported at 300 K and 30 K[32], respectively. As a result, future improvement in BP thin film material quality is likely to result in much enhanced mobilities. The temperature dependence of the carrier mobility in BP shows a characteristic behavior of 2D materials. The initial increase in carrier mobility from 10 to 100 K is most likely due to the dominating Coulomb scattering mechanism[41-42]. As temperature further increases, the mobility decreases due to phonon scattering following the power law relation $\sim T^{-\gamma}$ with $\gamma$ close to 0.5. This power law dependence is apparently weaker than in other 2D materials like graphene and TMDCs [43]. Further theoretical and experimental works are needed to clarify the temperature dependence of the mobility.

5. **Transistors using few-layer BP as channel material**

We also fabricated a series of transistors with a channel length of 1 μm using BP thin films with thickness varying from 2 to 15 nanometers. Fig. 5a and 5b denote the transfer and output characteristics of a BP transistor with a thickness of around 5 nm at room temperature. An on-off current ratio exceeding $10^5$ is achieved. Also plotted in Fig. 5a are the transfer characteristics in linear scale, from which field effect mobility of 205 cm$^2$/Vs are inferred using $\mu_{FE} = \frac{dI_{DS}}{dV_{BG}} \frac{L}{C_{ox} W V_{DS}}$, where $I_{DS}$ is the source-drain current, $V_{BG}$ is the back gate bias, $V_{DS}$ is the source-drain bias, $C_{ox}$ is the gate capacitance, W is the width of the transistor, and L is the channel length. Here, we emphasize that the field effect mobility is strongly affected by the traps in the device the intrinsic band mobility can be much higher than 205 cm$^2$/Vs[44]. Compared with graphene transistors, the BP devices consistently show much improved current saturation behavior as shown in Fig. 5b. Finally, we plot the scaling of on-off current ratio as a function of the BP thickness in the inset of Fig. 5a. For BP thickness of around 2 nanometers, the on-off current ratio is as high as $5 \times 10^5$. The much improved transistor performance as compared to graphene devices indicates that the BP transistors can overcome the major shortcomings of graphene transistors in terms of on-off current ratio and current saturation. Combined with a much higher mobility than TMDCs, BP devices have great potential for applications in thin film radio-frequency electronics and logic circuits.

In summary, we introduce black phosphorus as a unique anisotropic material with narrow bandgap and high carrier mobility to the layered 2D material family. It bridges the space between the semi-metallic graphene and various large gap TMDCs, making BP thin films an ideal candidate for various near and mid-infrared optoelectronic applications such as photodetectors and modulators. Moreover, due to its relatively high carrier mobility and good current saturation properties, few-layer BP has great potential to replace graphene as a novel and more favorable 2D material for high performance RF and logic applications in thin film electronics. Most importantly, its unique angle dependent properties allow a new degree of freedom for designing conceptually new optoelectronic and electronic devices not possible using other 2D materials that can take advantage of the strong anisotropic nature of this intriguing materials.

**Notes**

During the preparation of this manuscript, we became aware of two experimental works studying the electronic properties of black phosphorus on arXiv.org (arXiv:1401.4117v1 and arXiv:1401.4133v1).

**Methods**

**Transistor and Hall bar fabrication:** the fabrication of our devices starts with the exfoliation of BP thin films from bulk BP crystals onto 300 nm $SiO_2$ on Si substrate, which has pre-patterned alignment grids, using the micro-mechanical cleavage technique. The thickness of the $SiO_2$ was selected to provide the optimal optical contrast for locating BP flakes relative to the alignment grids and for identifying their number of layers. The number of BP layers was then confirmed by atomic force microscopy (AFM) based on its thickness. The next step was to pattern the metal layer, which forms the electrodes directly in contact with BP, using Vistec 100 KV electron-beam lithography system based on poly(methyl methacrylate) (950k MW PMMA) resist. We then evaporated 1 nm Ti/20 nm Pd/30 nm Au followed by lift-off to form the contacts.

**AFM and Raman spectroscopy:** Atomic force microscopy (AFM) for identifying the thin film thickness was performed on a Digital Instruments/Veeco Dimension 3000 system. Angle-resolved Raman spectroscopy was performed using a 532 nm Nd:YAG laser in the LabRAM ARAMIS system.

**IR spectroscopy:** IR extinction spectroscopy was performed in the 800 $cm^{-1}$ to 6000 $cm^{-1}$ range using a Bruker Optics Fourier Transfer Infrared spectrometer (Vertex 70) integrated with a Hyperion 2000 microscope system. Incident light polarization was achieved using an IR polarizer.

**Electrical characterization:** All electrical characterizations were done using Agilent 4155C semiconductor parameter analyzer and a Lakeshore cryogenic probe station with micromanipulation probes.

**Figure Captions**

**Figure 1: Characterization of the thickness of black phosphorus (BP) flakes**

(a) Layered crystal structure of black phosphorus, showing two adjacent puckered sheets with linked phosphorus atoms. The thickness of each BP layer is around 5.3 Å. (b) An atomic force microscopy (AFM) image of black phosphorus flakes. Layer numbers from 2 to > 50 are identified based on their thickness.

**Figure 2: Angle-resolved optical and DC conductivity of BP thin film**

(a) Angle-resolved infrared extinction spectra when light is polarized along the six directions as shown in the inset. Inset: an optical micrograph of a BP flake with a thickness of around 30 nm. (b) DC conductivity and IR extinction measured along the same six directions on this BP flake and plotted in polar coordinates. The angle-resolved DC conductance (solid dots) and the angle-resolved extinction of the same flake at 2700 cm$^{-1}$ (hollow squares) are shown on the same polar plot. Dots and squares of the same colour correspond to the directions indicated by the arrows in the inset in (a). Inset: an optical image of this BP flake with 12 electrodes spaced at 30° apart. The crystalline orientation as determined by angle-resolved IR measurements and angle-resolved DC conductance measurements match exactly. In both (a) and (b), $\phi$ is the angle between the x-direction and the 0° reference.

**Figure 3: Angle-resolved Raman scattering**

Angle-resolved Raman spectra of BP layered thin film with 532 nm linearly-polarized laser excitation incident in the z-direction. Raman spectra along x- (grey), D- (red), and y-directions (blue) are shown. D-direction is along 45° angle relative to x- and y-directions, as shown in the bottom panel. All spectra show $A_g^2$, $B_{2g}$, and $A_g^1$ modes. The peak positions do not change as excitation light polarization varies. However, the relative intensities of these three modes change significantly with incident light polarization. The angle-resolved Raman spectra can be used to identify the crystalline orientation of BP thin films, which is especially useful for relatively small flakes due to the small laser spot size (~1-2 μm).

**Figure 4: Angle-resolved Hall mobility**

Hall mobilities measured along x ($\mu_x$) and y ($\mu_y$) directions for BP thin films with a thickness of 8 and 15 nm, respectively. $\mu_x$ is about 1.8 times as large as $\mu_y$ and thicker films consistently show larger mobility than that of thinner films.

**Figure 5: BP field effect transistors**

(a) Transfer characteristics of a field effect transistor with a channel length of 1 μm using a BP thin film of around 5 nm thick. $I_{DS}$ and $V_{BG}$ are source-drain current and back gate bias, respectively. Left and right axes represent linear and logarithmic scale, respectively. Inset: on-off current ratio as a function of BP thickness. (b) Output characteristics of the same transistor showing current saturation. $V_{DS}$ is the drain-source bias. All the transistors have channel length of 1 μm. Inset: a schematic view of the transistor.

# Figure 1. Characterization of BP thickness

(a)

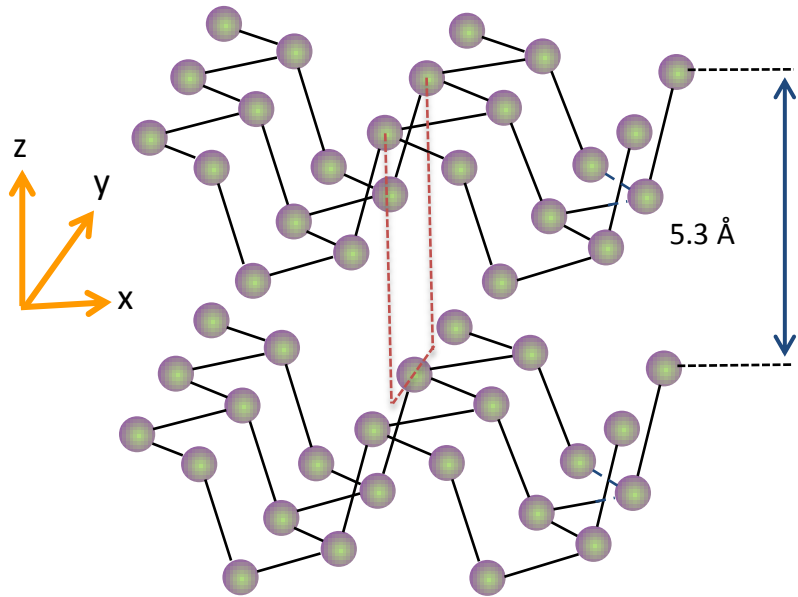

(b)

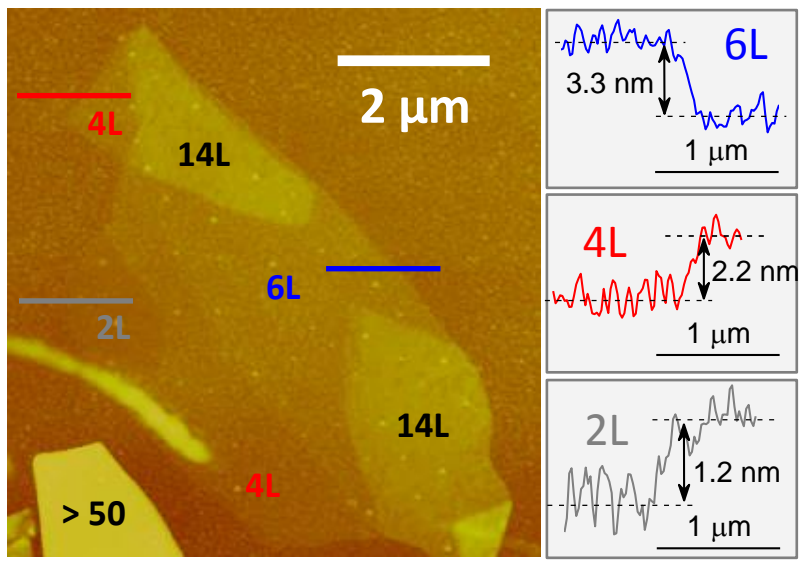

# Figure 2. Angle-resolved optical and dc conductivity

(a)
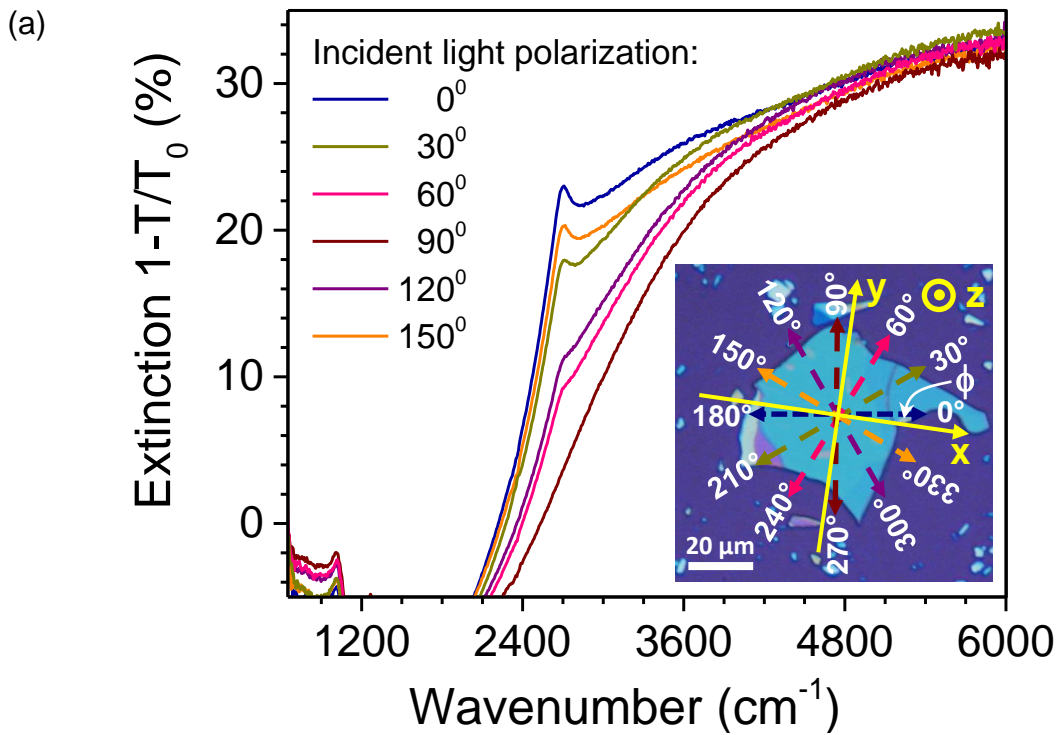

(b)
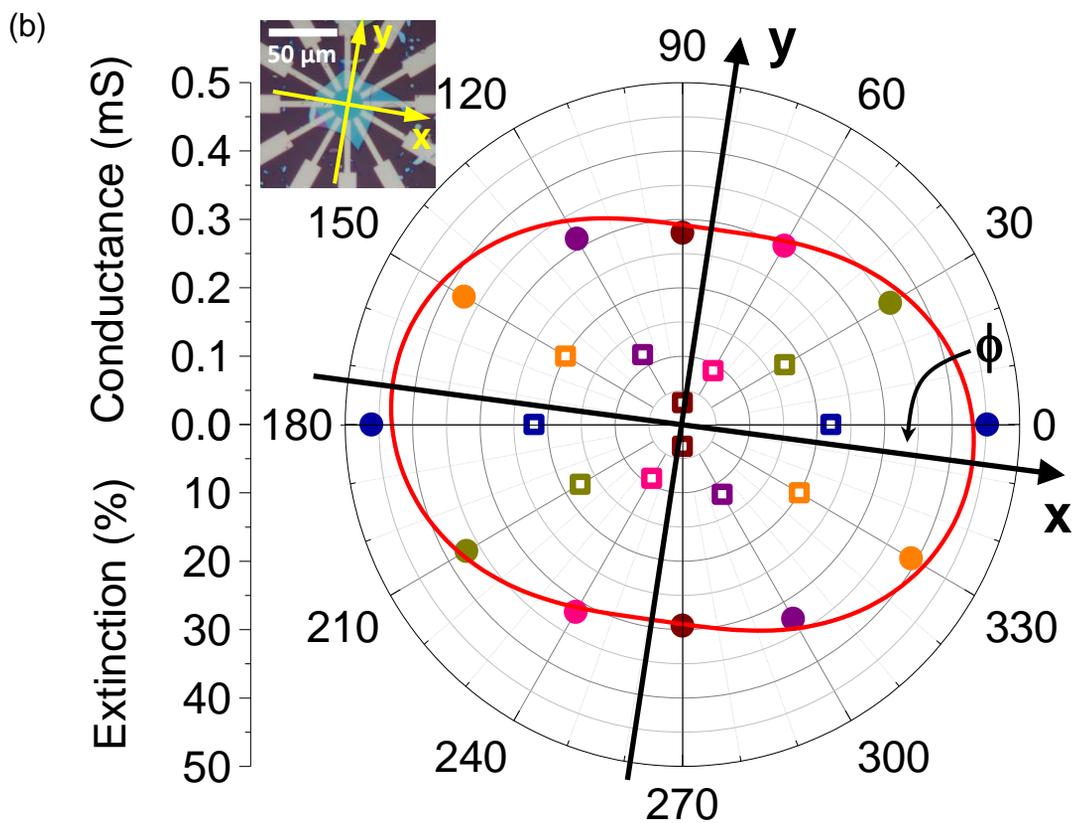

# Figure 3. Angle-resolved Raman measurements

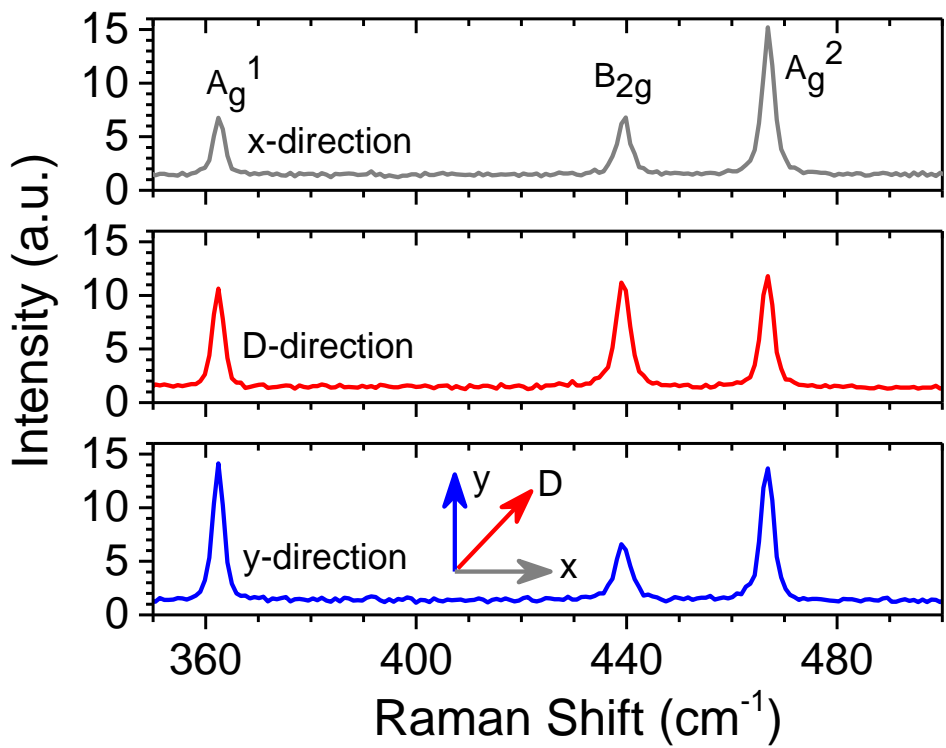

# Figure 4. Angle-resolved Hall mobility

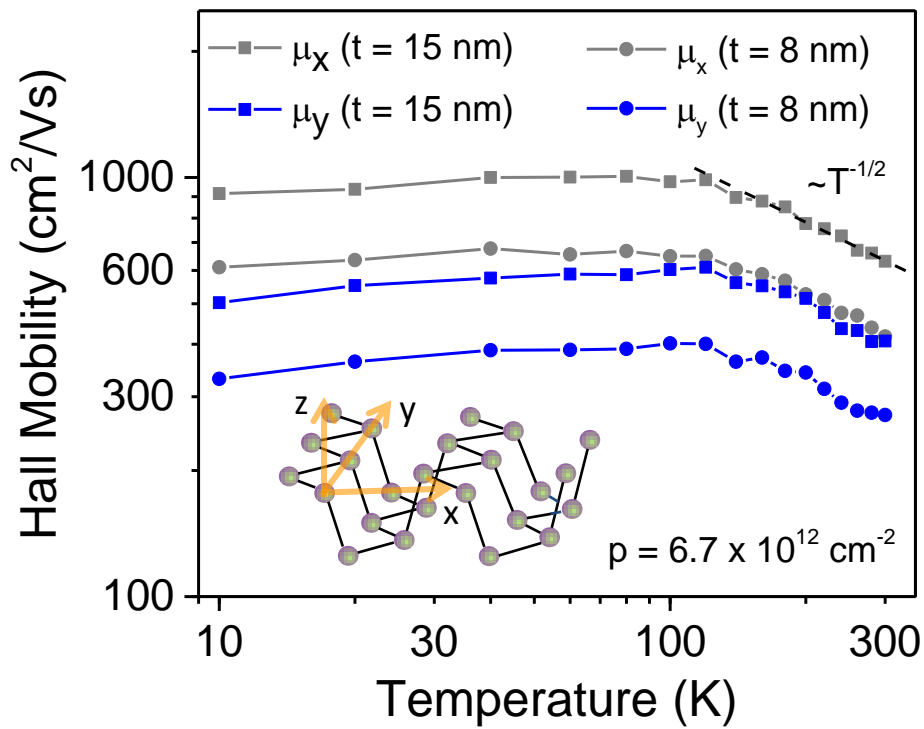

# Figure 5. Few-layer black phosphorus transistors

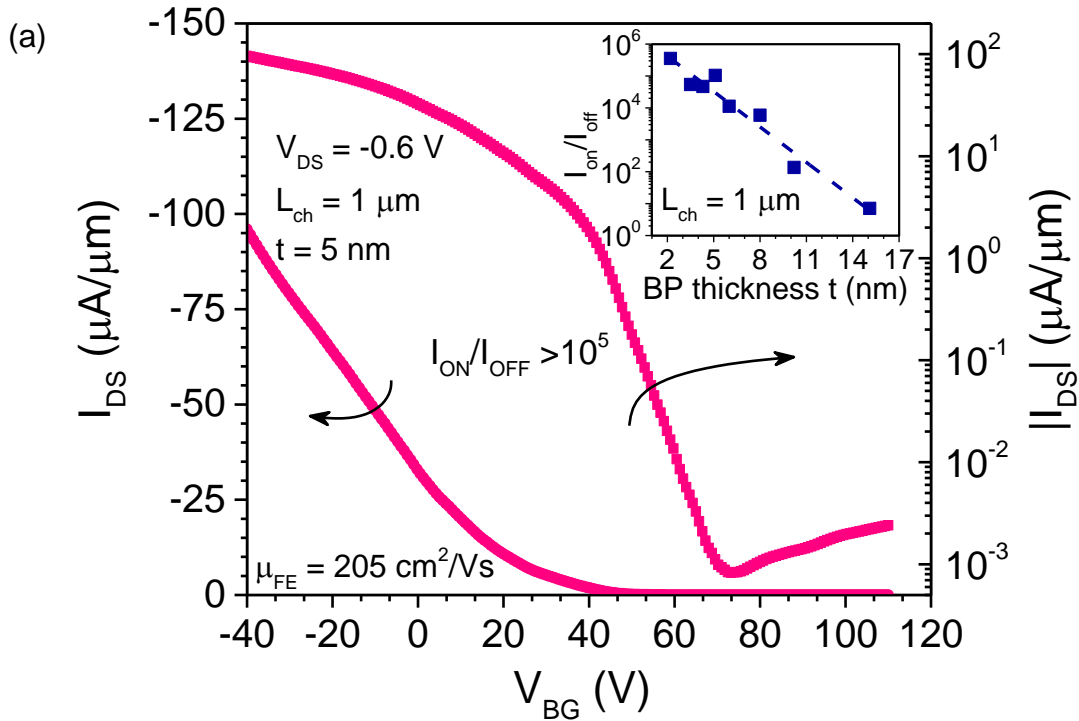

(a)

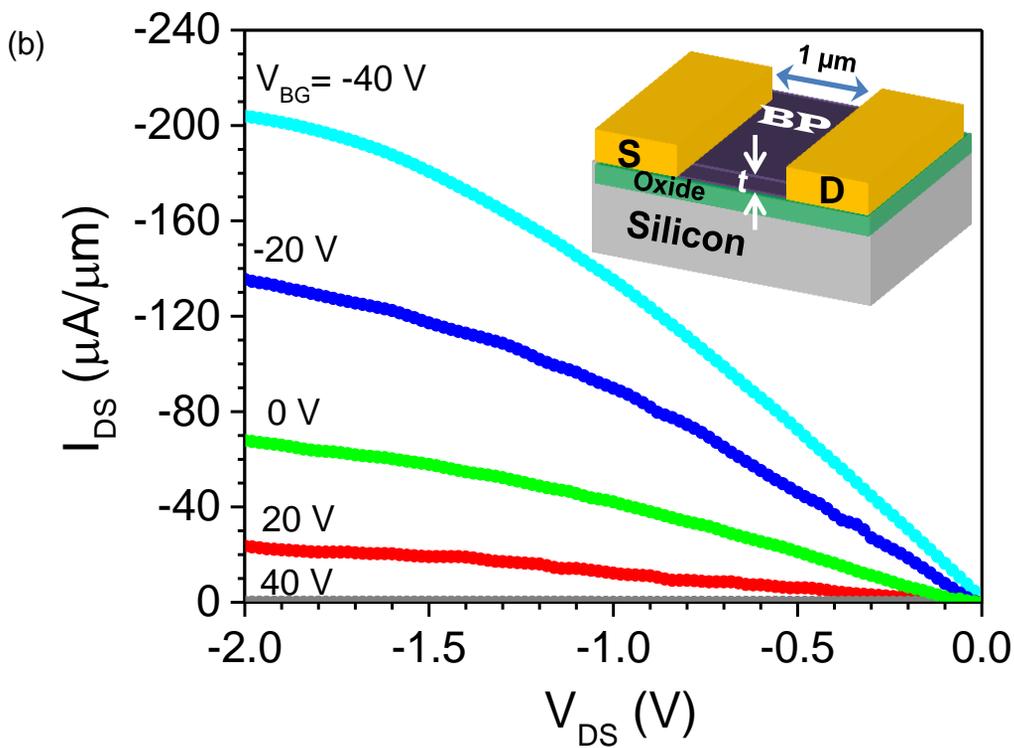

(b)